\documentstyle[11pt,paspconf,epsf]{article}

\begin{document}

\title{A WIYN Survey of Early-Type Barred Galaxies: Double Bars and
Central Structures}
\author{Peter Erwin and Linda S. Sparke}
\affil{Astronomy Department, University of Wisconsin--Madison, 475 N. Charter St., Madison, WI 53703}


\section{Introduction}

We present results from a preliminary analysis of a
recently-completed, multicolor imaging survey of nearby, early-type
barred galaxies in the field, carried out with the WIYN telescope and
supplemented with archival HST images.  This forms a reasonably
complete sample of nearby, bright, barred S0 and Sa galaxies in the
field.  The excellent seeing provided by WIYN allows us to examine the
galaxies for central features such as circumnuclear rings and
secondary bars; we find some evidence for dust lanes within secondary
bars.

The most striking results is the high frequency of double bars: our
analysis suggests that \textit{at least $\sim$20\% of barred S0--Sa
galaxies possess secondary bars}.  We also find an excellent candidate
{\it triply} barred galaxy: NGC 2681.

\section{The Sample and Observations}

Our sample was designed to address questions
about the frequency of multiple central structures in barred galaxies.
To minimize confusion due to dust, we concentrated on early type (S0
and Sa) galaxies.  Using the UGC, we chose all barred S0 and Sa
galaxies north of $-10^{\circ}$ meeting these criteria: $z \le 2000$
km/s, major axis $\ge 2$ arcmin, and axis ratio (major : minor) $\le
2$.  Galaxies in the Virgo Cluster were excluded, resulting in a total
of 38 galaxies.  Although there is a bias towards luminous, high
surface brightness galaxies, this is a reasonable complete set
of galaxies.

All but two of the galaxies were observed in the B and R bands between
December, 1995, and March, 1998, at the 3.5m WIYN telescope in Tucson,
Arizona.  Seeing ranged from 0.6--$1.3^{\prime\prime}$; the majority
of galaxies were observed in sub--arc second seeing.  Archival HST
data was also consulted for some of the galaxies.  For the two
galaxies not observed, excellent ground-based and HST data were
available in the literature.

\section{Analysis}

We search for secondary bars in our galaxies by fitting ellipses to
isophotes.  Bars are indicated by a peak in the ellipticity,
accompanied by a plateau or stationary point in the position angles of
the ellipses at roughly the same semi-major axis.  We also inspect the
B band images and color maps to ensure that dust lanes and star
formation do not create spurious bar-shaped isophotes.

\begin{figure}
\plotone{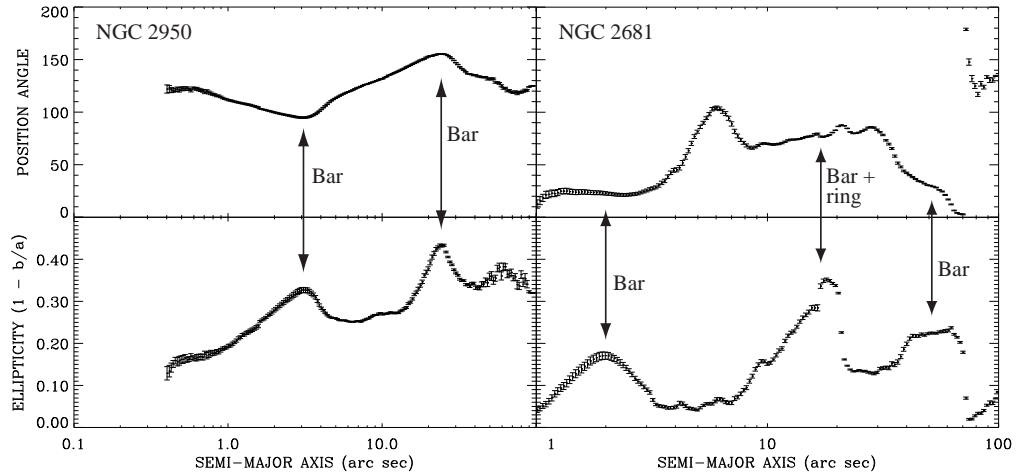}
\caption{Isophote fits for two galaxies from our sample.  NGC 2950 is
a typical double-barred galaxy; NGC 2681 is a \textit{triple}-barred
galaxy, with an inner ring around the middle bar.}

\end{figure}

\section{Preliminary Results}
Of the 22 galaxies we have analyzed so far, there are:

\begin{itemize}
\item 5 galaxies with clear secondary bars (NGC 2681, NGC 2859, NGC
2950, NGC 3945, NGC 4314 --- note that NGC 2681 has \textit{three}
bars!);

\item 5 more with possible secondary bars (NGC 2880, NGC 3185, NGC 3412, NGC
3941, NGC 4643);

\item 6 galaxies with \textit{no} secondary bar, and 6 galaxies too
dusty to determine their central stellar structure.
\end{itemize}

At least 26\% of our galaxies have clear or possible secondary bars.
We tentatively conclude that \textit{the fraction of barred S0 and Sa
galaxies in the field with secondary bars is at least 20\%.}

We can construct a similar sample of galaxies in the southern sky
(matching our sample criteria, but with $\delta \le -10^{\circ}$ and
excluding the Fornax cluster).  A search of the literature reveals 6
known double bars, out of the 28 galaxies in the sample.  The minimum
fraction of double bars in that sample is thus 21\%.  This strengthens
our belief that our (northern) sample is not unusual in its frequency of
double bars.  \textit{Double bars are surprisingly common in field S0
and Sa galaxies.}

\end{document}